\documentclass[elsart12,epsf,eqsecnum,graphics,cite]{revtex4}
\def\beq{\begin{equation}}

\newcommand{\del}{\partial}

\def\eeq{\end{equation}}
\usepackage{amsmath,amsfonts,amssymb,amsthm,nccmath,latexsym,mathtools}
\usepackage[mathscr]{euscript}
\usepackage{epsfig,epsf}
\usepackage{graphicx}
\usepackage{grffile}
\usepackage{xcolor}

\usepackage[hidelinks]{hyperref}
\hypersetup{
    colorlinks,
    citecolor=blue,
    filecolor=black,
    linkcolor=black,
    urlcolor=black
}
\hypersetup{linktocpage}

\usepackage{cancel}

\begin{document}


\voffset1.5cm

\title{On Entanglement Entropy of Maxwell fields in 3+1 dimensions with a slab geometry}
\author{ Candost Akkaya and Alex Kovner}

\affiliation{
Physics Department, University of Connecticut, 2152 Hillside
Road, Storrs, CT 06269-3046, USA}

\begin{abstract}
We calculate the entanglement entropy of a slab of finite width in the pure Maxwell theory. We find that a large part of entropy is contributed by the entanglement of a mode, nonlocal in terms of the transverse magnetic field degrees of freedom. Even though the entangled mode is nonlocal, its contribution to the entropy is local in the sense that the entropy of a slab of a finite thickness is equal to the entropy of the boundary plus a correction exponential in thickness of the slab.
\end{abstract}
\maketitle
\section{Introduction}
The entanglement properties of the vacuum of gauge theories is a very interesting subject. In particular entanglement entropy between two regions of space in quantum field theories has been a focus of many investigations triggered by the discovery of topological entropy in the context of quantum information theory\cite{Kitaev:2005dm},\cite{Levin:2006zz}. It can provide complementary information to standard correlation properties, in particular with theories with global degrees of freedom and a probe of possible long range dynamics.
 
 Technically such a calculation provides a challenge even in simple abelian theories.
 
 A calculation of von Neumann entropy is a complicated endeavor \cite{Witten:2018lha} and to date it has been performed either in conformal field theories using CFT methods \cite{Solodukhin:2008dh}, or in free field theories \cite{Casini:2009sr}. Even in free theories this calculation is not entirely straightforward. In particular there is no consensus to date on the result for entropy in abelian gauge theories \cite{Casini:2013rba,Casini:2014aia,Huang:2014pfa,Huerta:2018xvl}. The early calculation using Euclidean formulation found a nonstandard contact term \cite{Kabat:1995eq} whose existence is still controversial \cite{Donnelly:2015hxa}. In 2+1 dimensions pure Maxwell theory the calculation can be performed essentially using the equivalence of the theory of a free photon to that of a single massless scalar \cite{Agarwal:2016cir,Murciano:2020vgh}. Due to the fact that the bosonization techniques available to low dimensions do not generalize to higher dimensions, the separation of locally physical degrees of freedom in a gauge invariant way turns into a nontrivial matter. \cite{Donnelly:2011hn,Radicevic:2015sza}.
 
 In the previous paper \cite{first} we have calculated the entropy of entanglement between two halves of space in pure Maxwell theory, using a straightforward approach of integrating part of degrees of freedom in the vacuum wave function. A nontrivial aspect of this calculation is the necessity to choose gauge invariant physical degrees of freedom in order to reduce the Hilbert space. We chose to use the two components of magnetic field, parallel to the plane separating the two halves of space (transverse components) as our physical basis. This required us to solve the no-monopole condition in order to express the longitudinal component of the magnetic field in terms of the unconstrained $B_i$. 
 
 The entanglement entropy turned out to be  proportional to the area of the transverse plain with finite entropy density (for a given transverse momentum mode). An interesting feature of the calculation however, is that a final fraction of  the entanglement entropy  is contributed by a mode which is nonlocal in the longitudinal direction. The question to ask then is whether this signals a genuinely nonlocal long range entanglement, or whether the apparent nonlocality is due to the fact that the longitudinal magnetic field (after solving the no-monopole condition) is a nonlocal function of the transverse magnetic field components \footnote{We use the notions of "transverse" and "longitudinal" in the two dimensional sense, i.e. in the situation where the plane separating the two spatial regions is perpendicular to the $x_3$ axis, we refer to $B_{1,2}$ as transverse and to $B_3$ as longitudinal.}.
 
 To address this question in the current note we extend the calculation in \cite{first} to a situation that has more structure. We consider a bipartite system that consists of a slab of a finite width $d$ and  infinite transverse dimension and its complement consisting of two half spaces $R$ and $L$, to the right and the left of the slab respectively. We integrate out the degrees of freedom in the slab and calculate the reduced density matrix and associated entanglement entropy for the remaining system $L\oplus R$.
 
  If the theory exhibits genuine long range entanglement we would expect that due to the integration out of the fields inside the slab, the magnetic fields in $L$ should become strongly entangled with the magnetic fields in $R$ and this entanglement should be reflected in the entanglement entropy. Our results do not support this expectation. We find that the entanglement entropy does not exhibit such long range features, and instead approaches direct sum of the entropy in $L$ and in $R$. In a local theory we expect that the correction to the sum of entanglement entropies in $L$ and $R$ should decrease with the thickness of the slab as  $\Delta S_e\propto e^{-knd}$, where $k$ is the transverse momentum of the field mode in question, and $d$ - the thickness of the slab and $n$ is some integer. We indeed find such a correction with $n=2$ to the contribution of the entropy due to entanglement of the transverse mode. For the longitudinal mode we also find an exponentially suppressed correction. Interestingly, this contribution is suppressed by and additional factor $d/L$, where $L$ is the linear size of the system, $\Delta S\propto \frac{d}{L}e^{-2kd}$. We suspect that this apparent "super locality"  is due to our omission of the mixing of this mode with additional modes localized at the boundary
  of the slab. 
  
  We conclude that the apparent nonlocality arising in the calculation is merely due to the fact that the longitudinal mode of magnetic field is expressed nonlocally in terms of the transverse components. Nevertheless since the longitudinal component itself is a local field, the physical effect of entanglement  is local.

\section{Generalities}
We consider free Maxwell fields in 3+1 dimensions. The vacuum  wave functional of the theory of a free photon can be written in terms of magnetic field

\begin{align}\label{vacwf}
\langle A|\psi\rangle=\psi_0[\vec{A}]=Nexp\Big\{-\frac{1}{(2\pi)^2}\int d^3xd^3y\frac{B_i(x)B_i(y)}{|x-y|^2}\Big\} \quad\quad i=1,2,3
\end{align}

To define an unconstrained set of degrees of freedom we need to solve  the 
``no monopole'' condition
\begin{align}
\del_iB_i=0,\ \ \ \ i=1,2,3
\end{align}
We do this in the same way as in \cite{first}, namely by eliminating the third component of the magnetic field
\begin{align}
B_3(x,z)=B_3(x,0)-\int_{0}^z dz' \del_iB_i (x,z')\quad\quad i=1,2
\end{align}
where $x$ stands for the transverse coordinates, and from now on we use $i$ to denote transverse indexes only.
We find it convenient to separate the magnetic field $B_i$ into (two dimensionally) transverse and longitudinal components 
\begin{equation}
\partial_iB_i(x,z)=\chi(x,z); \ \ \ \ \ \epsilon_{ij}\partial_iB_j(x,z)=\zeta(x,z)
\end{equation}

Defining
\begin{equation}
\phi(x)\equiv B_3( x,0); 
\end{equation}
we have (where now by $x$ and $z$ we denote the transverse and longitudinal coordinates respectively)
\begin{align}\label{b3}
B_3(\bm{x},z)=\phi(x)-\int_0^{z} dz'\chi(x,z') 
\end{align}
The integration measure for the functional integral over the magnetic field is thus the Cartesian measure for the (planar) magnetic field $B_i(x,z)$ or equivalently $\chi$ and $\zeta$, and the planar field $\phi(x)$ which is defined as $\phi(x)=B_3(x,z=0)$.

The vacuum wave functional $\psi$ has the product form
\begin{equation} \psi[\zeta,\chi,\phi]=\psi[\zeta]\psi[\chi,\phi]
\end{equation}
The calculation therefore can be performed separately in the transverse and longitudinal sectors.

Consider the following geometry depicted on Fig. 1.

\begin{figure}[h!] \label{decomp}
  \includegraphics[scale=0.8]{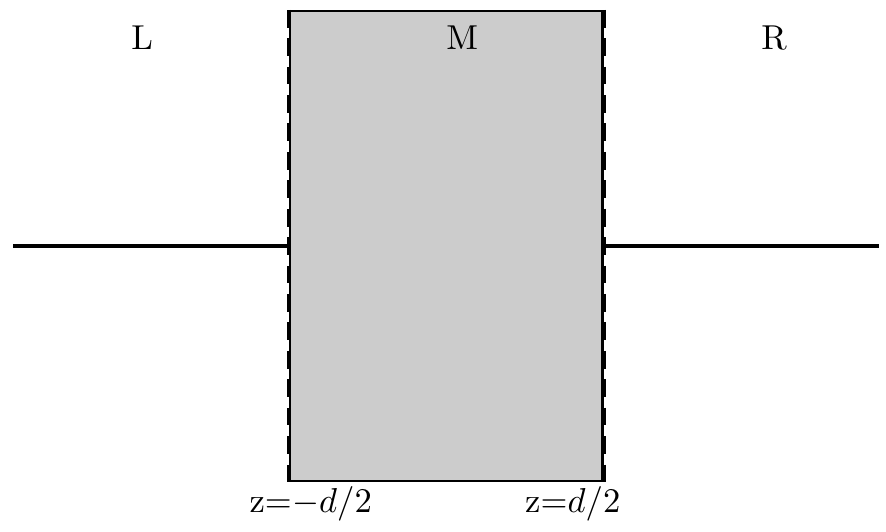}
  \caption{Decomposition of the space into three pieces denoted by L,M, and R.}
 
\end{figure}

Our goal is to integrate all physical degrees of freedom in side the slab, $-d/2<z<d/2$ and calculate the von Neumann entropy of the resulting reduced density matrix on the Hilbert space of degrees of freedom living in $L\oplus R$.

Note that since the geometry preserves translational invariance in the transverse direction, we can decompose the fields into transverse momentum modes which decouple in the wave function. We will use this Fourier representation and will perform the calculation for a given transverse momentum mode ${\bf k}$.
We decompose the fields as 

\begin{align}
\zeta(k,z)=\zeta^L(k,z)\Theta(-z-d/2)+\zeta^M(k,z)\Theta(z+d/2)\Theta(d/2-z)+\zeta^R(k,z)\Theta(z-d/2)\\
\chi(k,z)=\chi^L(k,z)\Theta(-z-d/2)+\chi^M(k,z)\Theta(z+d/2)\Theta(d/2-z)+\chi^R(k,z)\Theta(z-d/2)
\end{align}
We will reduce the vacuum density matrix over $\zeta^M$, $\chi^M$ and $\phi$, and calculate the von Neumann entropy of the reduced density matrix. Note that integrating over these variables is equivalent to integrating over both $B_i$ and $B_3$, and thus all local degrees of freedom inside the slab.

\section{The transverse sector.}

 The factor in the wave function that depends only on transverse fields is given by 

\begin{align}
\psi[\zeta]=N_\zeta exp \Bigg\{-\int dzdz' \zeta(k,z)\frac{K_0(k(z-z'))}{k^2}\zeta(-k,z') \Bigg\}
\end{align}
where $K_0$ is the Bessel function.
The partial trace of the density matrix for a given transverse momentum $k$  is given by

\begin{align}\label{dens}
\begin{split}
\rho[\zeta^L,\zeta^R,\zeta^{L'},\zeta^{R'}]&=N\int D\zeta^M\psi[\zeta^L,\zeta^R,\zeta^M]\psi^\dag[\zeta^{L'},\zeta^{R'},\zeta^M]\\
&=N\int D \zeta^M exp\Bigg\{-\Bigg[ \int_{LL}. \left[\zeta^L(k,z)\frac{K_0[k(z-z')]}{k^2}\zeta^{L}(-k,z')+\zeta^{L'}(k,z)\frac{K_0[k(z-z')]}{k^2}\zeta^{L'}(-k,z')\right]\\
&+\int_{RR}\left[\zeta^R(k,z)\frac{K_0[k(z-z')]}{k^2}\zeta^{R}(-k,z')+\zeta^{R'}(k,z)\frac{K_0[k(z-z')]}{k^2}\zeta^{R'}(-k,z')\right]\\
&+2\int_{RL}\left[\zeta^R(k,z)\frac{K_0[k(z-z')]}{k^2}\zeta^{L}(-k,z')+\zeta^{R'}(k,z)\frac{K_0[k(z-z')]}{k^2}\zeta^{L'}(-k,z')\right]\\
&+2\int_{MM}\zeta^M(k,z)\frac{K_0[k(z-z')]}{k^2}\zeta^{M}(-k,z')\\
&+2\int_{ML\{R\}}\zeta^M(k,z)\frac{K_0[k(z-z')]}{k^2}\left[\zeta^{L}(-k,z')+\zeta^{L'}(-k,z')+\zeta^{R}(-k,z')+\zeta^{R'}(-k,z')\right]   \Bigg]\Bigg\}
\end{split}
\end{align}
where we have introduced the abbreviations $\int_M=\int^{d/2}_{-d/2}$, $\int_L=\int^{-d/2}_{-\infty}$, $\int_R=\int^{\infty}_{d/2}$.

The Gaussian integration over  $\zeta^M$ amounts to the substitution of the solution of the "classical equations of motion"

\begin{align}\label{eom}
\int_MK_0[k(z-z')]\zeta^M(k,z')=-\frac{1}{2}\Bigg(\int_LK_0[k(z-z')][\zeta^L(k,z')+\zeta^{L'}(k,z')]+\int_RK_0[k(z-z')][\zeta^R(k,z')+\zeta^{R'}(k,z')]\Bigg)
\end{align}

 For convenience let us define

\begin{align}
f_1(z)=\sqrt{\frac{1}{8\pi k(d/2+z)}}e^{-k(d/2+z)} \ \ \ \ \ \ f_2(z)=\sqrt{\frac{1}{8\pi k(d/2-z)}}e^{-k(d/2-z)} \ \ \ \ \ \ F_{ij}\equiv\int_M dz k f_i(z)f_j(z)
\end{align}

As in \cite{first} we will work in the local approximation, namely assume that the main contribution to physical quantities of interest arise from the longitudinal distances such that  $kz>1$. In this approximation the classical solution reads

\begin{align}
\zeta^M(k,z)=-f_1(z)[\zeta^L(k,-d/2)+\zeta^{L'}(k,-d/2)]-f_2(z)[\zeta^R(k,d/2)+\zeta^{R'}(k,d/2)]
\end{align}

After the substitution of (\refeq{eom}) into the density matrix (\refeq{dens}), in this local limit we get

\begin{align}\label{dens2}
\begin{split}
\rho=Nexp\Bigg\{
 -\frac{\pi}{k^3}\Bigg[ &\int_L \zeta^L(k,z)\zeta^L(k,z)
   +\int_L \zeta^{L'}(k,z)\zeta^{L'}(k,z)
   + \int_R \zeta^{R}(k,z)\zeta^{R}(k,z)
    +\int_R \zeta^{R'}(k,z)\zeta^{R'}(k,z) 
 \\&+\frac{2}{k}\frac{e^{-kd}}{\sqrt{2\pi kd}}\zeta^L\left(k,-\frac{d}{2}\right)\zeta^R\left(-k,\frac{d}{2}\right)+
 \frac{2}{k}\frac{e^{-kd}}{\sqrt{2\pi kd}}\zeta^{L'}\left(k,-\frac{d}{2}\right)\zeta^{R'}\left(-k,\frac{d}{2}\right)
 \\&-\frac{2}{k}F_{11}\left[\zeta^L\left(k,-\frac{d}{2}\right)+\zeta^{L'}\left(k,-\frac{d}{2}\right)\right]\left[\zeta^L\left(k,-\frac{d}{2}\right)+\zeta^{L'}\left(k,-\frac{d}{2}\right)\right]
  \\&-\frac{2}{k}F_{22}\left[\zeta^R\left(k,\frac{d}{2}\right)+\zeta^{R'}\left(k,\frac{d}{2}\right)\right]\left[\zeta^R\left(k,\frac{d}{2}\right)+\zeta^{R'}\left(k,\frac{d}{2}\right)\right]
   \\&-\frac{4}{k}F_{12}\left[\zeta^L\left(k,-\frac{d}{2}\right)+\zeta^{L'}\left(k,-\frac{d}{2}\right)\right]\left[\zeta^R\left(k,\frac{d}{2}\right)+\zeta^{R'}\left(k,\frac{d}{2}\right)\right]
 \Bigg]  \Bigg\}
 \end{split}
\end{align}

It is clear from eq. (\refeq{dens2}) that the leading contribution to the entanglement is coming from the fields that are pinned at $\pm d/2$, since for any other mode the reduced density matrix is in the product form. 
Keeping only the modes $\zeta(k,z=\pm d/2)$ we have

\begin{align}\label{dens3}
\begin{split}
\rho=Nexp\Bigg\{
 -\frac{\pi}{k^4}\Bigg[ & \zeta^L(k,-\frac{d}{2})\zeta^L(k,-\frac{d}{2})
   + \zeta^{L'}(k,-\frac{d}{2})\zeta^{L'}(k,-\frac{d}{2})
   +  \zeta^{R}(k,\frac{d}{2})\zeta^{R}(k,\frac{d}{2})
    +\zeta^{R'}(k,\frac{d}{2})\zeta^{R'}(k,\frac{d}{2}) 
 \\&+2\frac{e^{-kd}}{\sqrt{2\pi kd}}\zeta^L\left(k,-\frac{d}{2}\right)\zeta^R\left(-k,\frac{d}{2}\right)+
 2\frac{e^{-kd}}{\sqrt{2\pi kd}}\zeta^{L'}\left(k,-\frac{d}{2}\right)\zeta^{R'}\left(-k,\frac{d}{2}\right)
 \\&-2F_{11}\left[\zeta^L\left(k,-\frac{d}{2}\right)+\zeta^{L'}\left(k,-\frac{d}{2}\right)\right]\left[\zeta^L\left(k,-\frac{d}{2}\right)+\zeta^{L'}\left(k,-\frac{d}{2}\right)\right]
  \\&-2F_{22}\left[\zeta^R\left(k,\frac{d}{2}\right)+\zeta^{R'}\left(k,\frac{d}{2}\right)\right]\left[\zeta^R\left(k,\frac{d}{2}\right)+\zeta^{R'}\left(k,\frac{d}{2}\right)\right]
   \\&-4F_{12}\left[\zeta^L\left(k,-\frac{d}{2}\right)+\zeta^{L'}\left(k,-\frac{d}{2}\right)\right]\left[\zeta^R\left(k,\frac{d}{2}\right)+\zeta^{R'}\left(k,\frac{d}{2}\right)\right]
 \Bigg]  \Bigg\}
 \end{split}
\end{align}
We have discretized the longitudinal direction with the UV "cutoff" $k$ consistently with the local approximation.
The density matrix (\refeq{dens3}) is of the form

\begin{align}\label{dens4}
\rho=N exp\left\{ - \left[\alpha(X-F)\alpha +\alpha'(X-F)\alpha' -2\alpha F\alpha'        \right]  \right\}
\end{align}

where
\begin{align}
\alpha=\begin{pmatrix} \zeta^L(k,-d/2) \\ \zeta^R(k,d/2)\end{pmatrix} 
\ \ \ \ \ \ 
X=\frac{\pi}{k^4}\begin{pmatrix} 1&e^{-kd}/\sqrt{2\pi kd}\\e^{-kd}/\sqrt{2\pi kd}&1\end{pmatrix} 
\ \ \ \ \ \ 
F=\frac{\pi}{k^4}\begin{pmatrix} \gamma&e^{-kd}/4\\e^{-kd}/4&\gamma\end{pmatrix}
\end{align}
with $\gamma$ a pure number 
\begin{align}
\gamma=
\frac{1}{4\pi^2}\int_0^\infty dx K_0^2(x)\approx 0.0625
\end{align}

For the density matrix (\refeq{dens4}) the entanglement entropy reads \cite{Kovner2015a} 

\begin{align}
S_{\zeta}=\frac{1}{2}tr \left[ log\left\{\frac{U-1}{4}\right\} +\sqrt{U}\text{cosh}^{-1}\left\{\frac{U+1}{U-1}\right\}\right] \ \ \ \ \ U=(X-2F)^{-1}X
\end{align}
To understand better this expression we consider the regime of  large slab width, i.e.  $kd\gg 1$. This obviously can also be interpreted as high momentum regime. Physically one expects that the main contribution to entanglement entropy comes from these sort of momenta, since for momenta lower than the inverse slab width we expect the wave function to be practically unaffected by integrating ou the fields inside the slab. Thus no entanglement entropy should be associated with momentum modes with $kd\ll 1$.  

For large slab the off diagonal terms in $X$ are suppressed by a power of $kd$ relative to those in $F$, and can be neglected. Expanding to second order in $e^{-kd}$, the eigenvalues of the matrix $U$ are

\begin{align}\label{eigenvalues}
\lambda_\mp=\lambda\left(1\mp \frac{1}{2}\lambda e^{-kd}+\frac{1}{4}\lambda^2e^{-2kd}\right)\ \ \ \ \  \lambda=\frac{1}{1-2\gamma} \ \ \ 
\end{align}
We need to keep second order terms in $e^{-kd}$ , since as we will see, the first order terms cancel  in the expression for entropy.
With this we get

\begin{align}
S=\left(\log \left(\frac{\lambda -1}{4}\right)+\sqrt{\lambda } \cosh ^{-1}\left(\frac{\lambda +1}{\lambda
   -1}\right)\right)+\frac{1}{32} e^{-2kd} \left(3 \lambda ^{5/2} \cosh ^{-1}\left(\frac{\lambda +1}{\lambda -1}\right)-\frac{2 \lambda
   ^3}{\lambda -1}\right)
\end{align}






This expression exhibits the expected properties. The leading term at large $kd$ is equal to twice the result of \cite{first} where we have calculated the entropy due to one boundary by integrating magnetic fields in half space. 
For a local theory we indeed expect that each boundary will contribute to the entropy independently if they are far apart. The first correction is exponentially suppressed in $kd$ indicating that momentum $k$ plays the role of the inverse length over which the transverse degrees of freedom ($\zeta$) are entangled in the longitudinal direction. 
We also note that the correction to the leading term in entropy due to the finite width of the slab is negative. This is again expected in the local theory since for a finite slab width there is a finite amount of coherence left in the density matrix between the right and left half spaces, and thus the entropy is lower than in the $d\rightarrow \infty$ limit.
Those are all expected properties in the transverse sector,  as the apparent nonlocality observed in \cite{first} pertains to the longitudinal sector only. 

It is nevertheless interesting to note, that although the longitudinal scale that appears in the solutions to the "classical equations of motion" is the inverse of the momentum $k$, the final result for the entanglement entropy does not contain terms of order $e^{-kd}$. Although such terms are present in the eigenvalues of the matrix $U$, eq.(\ref{eigenvalues}), they cancel in the expression for the entanglement entropy, and the leading correction is of order $e^{-2kd}$. This is natural if the entropy depends not on the field $B$, but only on its square $B^2$,

We now turn to the calculation in the longitudinal sector.







\section{The Longitudinal sector.}

 

Let us now consider the density matrix for the longitudinal fields. 


The longitudinal field dependent factor in the wave function is 

\begin{align}
\psi_\chi=N_\chi  \ exp\left\{-\int dz dz' \int d^2 k \left\{\chi(k,z)\frac{K_0[k(z-z')]}{k^2}\chi(-k,z')
+  \left[\phi-\int_0^z du\chi(k,u)\right]  K_0[k(z-z')]   \left[\phi-\int_0^{z'} dv\chi(-k,v)\right]  \right\}  
\right\}
\end{align}
or in terms of our field decomposition
\begin{eqnarray}\label{psichi}
\psi_\chi&=&N_\chi  \ exp\Bigg\{-2\pi \frac{L}{k}\phi(k)\phi(-k)\\
&&+\frac{2\pi}{k}\phi(k)\Big[\int_0^{d/2} (L-z)\chi_M(-k,z)-\int_{-d/2}^0(L+z)\chi_M(-k,z)+\int_{d/2}^L(L-z)\chi_R(-k,z)-\int_{-L}^{-{d/2}}(L+z)\chi_L(-k.z)\Big]\nonumber\\
&&-\int_{RR}\chi_R(k,z)\left[\frac{1}{k^2}K_0(k(z-z'))+\int_{u>z;\ v>z'}K_0(k(u-v))\right]\chi_R(-k,z')\nonumber\\
&&-\int_{LL}\chi_L(k,z)\left[\frac{1}{k^2}K_0(k(z-z'))+\int_{u<z;\ v<z'}K_0(k(u-v))\right]\chi_L(-k,z')\nonumber\\
&&-2\int_{LR}\chi_L(k,z)\left[\frac{1}{k^2}K_0(k(z-z'))-\int_{u<z; v>z'}K_0(u-v)\right]\chi_R(-k,z')\nonumber\\
&&-\int_{z>0,z'>0}\chi_M(k,z)\left[\frac{1}{k^2}K_0(k(z-z'))+\int_{u>z;\ v>z'}K_0(k(u-v))\right]\chi_M(-k,z')\nonumber\\
&&-\int_{z<0,z'<0}\chi_M(k,z)\left[\frac{1}{k^2}K_0(k(z-z'))+\int_{u<z;\ v<z'}K_0(k(u-v))\right]\chi_M(-k,z')\nonumber\\
&&-2\int_{z>0,z'<0}\chi_M(k,z)\left[\frac{1}{k^2}K_0(k(z-z'))-\int_{u>z;\ v<z'}K_0(k(u-v))\right]\chi_M(-k,z')\nonumber\\
&&-2\int_{z>0,R}\chi_M(k,z)\left[\frac{1}{k^2}K_0(k(z-z'))+\int_{u>z;\ v>z'}K_0(k(u-v))\right]\chi_R(-k,z')\nonumber\\
&&-2\int_{z<0,L}\chi_M(k,z)\left[\frac{1}{k^2}K_0(k(z-z'))+\int_{u<z;\ v<z'}K_0(k(u-v))\right]\chi_L(-k,z')\nonumber\\
&&-2\int_{z>0,L}\chi_M(k,z)\left[\frac{1}{k^2}K_0(k(z-z'))-\int_{u>z;\ v<z'}K_0(k(u-v))\right]\chi_L(-k,z')\nonumber\\
&&-2\int_{z<0,R}\chi_M(k,z)\left[\frac{1}{k^2}K_0(k(z-z'))-\int_{u<z;\ v>z'}K_0(k(u-v))\right]\chi_R(-k,z')\Bigg\}\nonumber
\end{eqnarray}
where we have introduced an infrared cutoff $L$ regulating the longitudinal extent of space.

In the local approximation we can use the asymptotic form of the Bessel function. In this approximation we  have 
\begin{equation}
\int_{u>z; \ v<z'}K_0(k(u-v))=\frac{1}{k^2}K_0(k(z-z')); \ \ \  {\rm for} \ \ z>z'
\end{equation}
We also have
\begin{eqnarray}\label{relations}
\int_{u>z; \ v>z'}K_0(k(u-v))&= &  \int_{u>z; \ v}K_0(k(u-v))-\int_{u>z; \ v<z'}K_0(k(u-v))=\frac{\pi}{k}(L-z)-     \frac{1}{k^2}K_0(k(z-z')); \ {\rm for } \ z>z' \nonumber\\
\int_{u<z; \ v<z'}K_0(k(u-v))&= &  \int_{u; \ v<z'}K_0(k(u-v))-\int_{u>z; \ v<z'}K_0(k(u-v))=\frac{\pi}{k}(z'+L)-     \frac{1}{k^2}K_0(k(z-z')); \  {\rm for}\  z<z'
\end{eqnarray}
The relations eq.(\ref{relations}) are only valid as long as $|z-z'|\gg k^{-1}$. We therefore have to be careful using them especially if the integration over $z$ and $z'$ contains the vicinity of the point $z=z'$. Close to this point we should not use eq.(\ref{relations}), but instead we can neglect the terms of the form $\int_{u,v} K_0(k(u-v))$ relative to the unintegrated Bessel function in eq.(\ref{psichi}). The latter ones  with our resolution should be approximated by the delta function, $K(x)\approx \pi\delta(x)$. With this in mind we can simplify the wave function as follows

\begin{eqnarray}
\psi_\chi&=&N_\chi  \ exp\Bigg\{-2\pi \frac{L}{k}\phi(k)\phi(-k)\\
&&+\frac{2\pi}{k}\phi(k)\Big[\int_0^{d/2} (L-z)\chi_M(-k,z)-\int_{-d/2}^0(L+z)\chi_M(-k,z)+\int_{d/2}^L(L-z)\chi_R(-k,z)-\int_{-L}^{-d/2}(L+z)\chi_L(-k.z)\Big]\nonumber\\
&&-\frac{\pi}{k^3}\int_z \chi_R(k,z)]\chi_R(-k,z)-\frac{2\pi}{k}\int_{RR; z>z'}\chi_R(k,z)(L-z)\chi_R(-k,z')\nonumber\\
&&-\frac{\pi}{k^3}\int_z \chi_L(k,z)]\chi_L(-k,z)-\frac{2\pi}{k}\int_{LL; z>z'}\chi_L(k,z)(L+z')\chi_L(-k,z')\nonumber\\
&&-\frac{\pi}{k^3}\int_z \chi_M(k,z)]\chi_M(-k,z)\nonumber\\
&&-\frac{2\pi}{k}\int_{z>z'>0}\chi_M(k,z)(L-z)\chi_M(-k,z')-\frac{2\pi}{k}\int_{z<z'<0}\chi_M(k,z)(L+z)\chi_M(-k,z')\nonumber\\
&&-\frac{2\pi}{k}\int_{z>0,R}\chi_M(k,z)(L-z') \chi_R(-k,z')
-\frac{2\pi}{k}\int_{z<0,L}\chi_M(k,z)(L+z')\chi_L(-k,z')\nonumber
\end{eqnarray}

To obtain the reduced density matrix, we trace over $\chi^M$  and $\phi$
\begin{eqnarray}\label{action}
\rho_\chi&=&N_\chi  \int D\chi_MD\phi\ exp\Bigg\{-4\pi \frac{L}{k}\phi(k)\phi(-k)\\
&&+\frac{4\pi}{k}\phi(k)\Big[\int_0^{d/2} (L-z)\chi_M(-k,z)-\int_{-d/2}^0(L+z)\chi_M(-k,z)\Big]\nonumber\\
&&+\frac{2\pi}{k}\phi(k)\Big[
\int_{d/2}^L(L-z)[\chi_R(-k,z)+\chi_R'(-k,z)]-\int_{-L}^{-d/2}(L+z)[\chi_L(-k.z)+\chi_L'(-k,z)]\Big]\nonumber\\
&&-\frac{\pi}{k^3}\int_z \chi_R(k,z)\chi_R(-k,z)-\frac{2\pi}{k}\int_{RR; z>z'}\chi_R(k,z)(L-z)\chi_R(-k,z')\nonumber\\
&&-\frac{\pi}{k^3}\int_z \chi_R'(k,z)\chi_R'(-k,z)-\frac{2\pi}{k}\int_{RR; z>z'}\chi_R'(k,z)(L-z)\chi_R'(-k,z')\nonumber\\
&&-\frac{\pi}{k^3}\int_z \chi_L(k,z)\chi_L(-k,z)-\frac{2\pi}{k}\int_{LL; z>z'}\chi_L(k,z)(L+z')\chi_L(-k,z')\nonumber\\
&&-\frac{\pi}{k^3}\int_z \chi_L'(k,z)\chi_L'(-k,z)-\frac{2\pi}{k}\int_{LL; z>z'}\chi_L'(k,z)(L+z')\chi_L'(-k,z')\nonumber\\
&&-\frac{2\pi}{k^3}\int_z \chi_M(k,z)\chi_M(-k,z)\nonumber\\
&&-\frac{4\pi}{k}\int_{z>z'>0}\chi_M(k,z)(L-z)\chi_M(-k,z')-\frac{4\pi}{k}\int_{z<z'<0}\chi_M(k,z)(L+z)\chi_M(-k,z')\nonumber\\
&&-\frac{2\pi}{k}\int_{z>0,R}\chi_M(k,z)(L-z') [\chi_R(-k,z')+\chi_R'(-k,z')]
-\frac{2\pi}{k}\int_{z<0,L}\chi_M(k,z)(L+z')[\chi_L(-k,z')+\chi_L'(-k,z')]\Bigg\}\nonumber
\end{eqnarray}
As before, we should solve the "classical equations of motion" that follow by differentiating the "action" in eq.(\ref{action}) with respect to $\chi_M$ and $\phi$
\begin{eqnarray}
&&
-\frac{4\pi}{k^3}\chi_M(k,z)-\frac{4\pi}{k}\theta(z)\left[(L-z)\int_0^z\chi_M(k,z')+\int_z^{d/2}(L-z')\chi_M(k,z')\right]\nonumber\\
&&-\frac{4\pi}{k}\theta(-z)\left[(L+z)\int_z^0\chi_M(k,z')+\int_{-d/2}^z(L+z')\chi_M(k,z')\right]\nonumber\\
&&-\frac{2\pi}{k}\theta(z)\int_R(L-z')[\chi_R(k,z')+\chi_R'(k,z')]-\frac{2\pi}{k}\theta(-z)\int_L(L+z')[\chi_L(k,z')+\chi_L'(k,z')]\nonumber\\
&&+\theta(z)\frac{4\pi}{k}(L-z)\phi(k)-\theta(-z)\frac{4\pi}{k}(L+z)\phi(k)=0\nonumber\\
&&2L\phi(k)=\Big[\int_0^{d/2} (L-z)\chi_M(-k,z)-\int_{-d/2}^0(L+z)\chi_M(-k,z)\Big]+\frac{1}{2}(c_R-c_L)
\end{eqnarray}
with
\begin{equation}
c_R\equiv 
\int_R(L-z)[\chi_R(-k,z)+\chi_R'(-k,z)]; \ \ c_L\equiv \int_L(L+z)[\chi_L(-k.z)+\chi_L'(-k,z)]
\end{equation}
We differentiate the first equation twice to obtain for $z\ne 0$
\begin{equation}
\frac{d^2}{dz^2}\chi_M(k,z)=k^2\chi_M(k,z)
\end{equation}
with the solution
\begin{equation}\label{sol1}
\chi_M(k,z)=\theta(z)[a_1e^{kz}+a_2e^{-kz}]+\theta(-z)[b_1e^{kz}+b_2e^{-kz}]
\end{equation}
The coefficients are determined by substituting this form of the solution  back into the equations of motion:
\begin{eqnarray}\label{sol2}
&&\phi(k)=\frac{1}{k}(a_2-a_1)=\frac{1}{k}(b_2-b_1)\\
&&a_2e^{-kd/2}(d/2-L+\frac{1}{k})-a_1e^{kd/2}(d/2-L-\frac{1}{k})=-\frac{k}{2}c_R\nonumber\\
&&b_1e^{-kd/2}(d/2-L+\frac{1}{k})-b_2e^{kd/2}(d/2-L-\frac{1}{k})=-\frac{k}{2}c_L\nonumber\\
&&a_1+a_2=b_1+b_2
\end{eqnarray}
Approximating $L-d/2-\frac{1}{k}\approx L-d/2$  etc. for large $kd$ we get
\begin{eqnarray}
&&a_1=b_1=\frac{k}{4(d/2-L)\sinh(kd)}\left[c_Le^{-kd/2}+c_Re^{kd/2}\right]\\
&&a_2=b_2=\frac{k}{4(d/2-L)\sinh(kd)}\left[c_Le^{kd/2}+c_Re^{-kd/2}\right]\nonumber\\
&&\phi=\frac{1}{4(d/2-L)\cosh(kd/2)}[c_L-c_R]\nonumber\\
&&\chi_M(z)=\frac{k}{2(d/2-L)\sinh(kd)}\left[c_L\cosh(k(z-d/2))+c_R\cosh(k(d/2+z))\right]
\end{eqnarray}

Finally, substituting the solution back into the "action" we obtain for the reduced density matrix 
\begin{eqnarray}
&&\rho_\chi[\chi_L,\chi_R;\chi_L',\chi_R']=N_\chi  \ exp\Bigg\{
-\frac{\pi}{k^3}\int_z \chi_R(k,z)\chi_R(-k,z)-\frac{2\pi}{k}\int_{RR; z>z'}\chi_R(k,z)(L-z)\chi_R(-k,z')\nonumber\\
&&-\frac{\pi}{k^3}\int_z \chi_R'(k,z)\chi_R'(-k,z)-\frac{2\pi}{k}\int_{RR; z>z'}\chi_R'(k,z)(L-z)\chi_R'(-k,z')\\
&&-\frac{\pi}{k^3}\int_z \chi_L(k,z)\chi_L(-k,z)-\frac{2\pi}{k}\int_{LL; z>z'}\chi_L(k,z)(L+z')\chi_L(-k,z')\nonumber\\
&&-\frac{\pi}{k^3}\int_z \chi_L'(k,z)\chi_L'(-k,z)-\frac{2\pi}{k}\int_{LL; z>z'}\chi_L'(k,z)(L+z')\chi_L'(-k,z')\nonumber\\
&&+\frac{\pi}{k}\phi(k)\Big[
\int_{d/2}^L(L-z)[\chi_R(-k,z)+\chi_R'(-k,z)]-\int_{-L}^{-d/2}(L+z)[\chi_L(-k.z)+\chi_L'(-k,z)]\Big]\nonumber\\
&&-\frac{\pi}{k}\int_{z>0,R}\chi_M(k,z)(L-z') [\chi_R(-k,z')+\chi_R'(-k,z')]
-\frac{\pi}{k}\int_{z<0,L}\chi_M(k,z)(L+z')[\chi_L(-k,z')+\chi_L'(-k,z')]\Bigg\}\nonumber
\end{eqnarray}
where now $\chi_M$ and $\phi$ are given by eqs.(\ref{sol1},\ref{sol2}).
This can be simplified
\begin{eqnarray}
&&\rho_\chi[\chi_L,\chi_R;\chi_L',\chi_R']=N_\chi  \ exp\Bigg\{
-\frac{\pi}{k^3}\int_z \chi_R(k,z)\chi_R(-k,z)-\frac{2\pi}{k}\int_{RR; z>z'}\chi_R(k,z)(L-z)\chi_R(-k,z')\nonumber\\
&&-\frac{\pi}{k^3}\int_z \chi_R'(k,z)\chi_R'(-k,z)-\frac{2\pi}{k}\int_{RR; z>z'}\chi_R'(k,z)(L-z)\chi_R'(-k,z')\\
&&-\frac{\pi}{k^3}\int_z \chi_L(k,z)\chi_L(-k,z)-\frac{2\pi}{k}\int_{LL; z>z'}\chi_L(k,z)(L+z')\chi_L(-k,z')\nonumber\\
&&-\frac{\pi}{k^3}\int_z \chi_L'(k,z)\chi_L'(-k,z)-\frac{2\pi}{k}\int_{LL; z>z'}\chi_L'(k,z)(L+z')\chi_L'(-k,z')\nonumber\\
&&+\frac{\pi}{k}\phi(k)\Big[c_R-c_L
\Big]-\frac{\pi}{k}\int_{z>0}\chi_M(k,z) c_R 
-\frac{\pi}{k}\int_{z<0}\chi_M(k,z)c_L \Bigg\}\nonumber
\end{eqnarray}

Calculating the relevant integrals 

\begin{eqnarray}
&\nonumber
\frac{\pi}{k}c_R\int_0^{d/2}\chi^M(z)
=-\frac{\pi c_R}{2kL sinh(kd)}\Bigg\{c_L \ sinh(k(z-d/2))+c_R \ sinh(k(z+d/2))\Bigg\}\Bigg|_0^{d/2}
\\&\nonumber
=-\frac{\pi c_R}{2kL sinh(kd)}\Bigg\{c_R \ sinh(kd)+sinh(kd/2)(c_L-c_R)\Bigg\}
\\&
\sim \boxed{-\frac{\pi}{2kL}c_R\left\{c_R\left(1-e^{-kd/2}\right)+c_Le^{-kd/2}\right\}}
\\&\nonumber
\frac{\pi}{k}c_L\int_{z<0}\chi^M(z)=-\frac{\pi c_L}{2kLsinh(kd)}\Bigg\{c_L \ sinh(kd)+sinh(kd/2)(c_R-c_L)\Bigg\}
\\&
\sim \boxed{-\frac{\pi}{2kL}c_L\left\{c_L\left(1-e^{-kd/2}\right)+c_Re^{-kd/2}\right\}}
\end{eqnarray}

we finally obtain
\begin{eqnarray}\label{rhos}
&&\rho_\chi[\chi_L,\chi_R;\chi_L',\chi_R']=N_\chi  \ exp\Bigg\{
-\frac{\pi}{k^3}\int_z \chi_R(k,z)\chi_R(-k,z)-\frac{2\pi}{k}\int_{RR; z>z'}\chi_R(k,z)(L-z)\chi_R(-k,z')\nonumber\\
&&-\frac{\pi}{k^3}\int_z \chi_R'(k,z)\chi_R'(-k,z)-\frac{2\pi}{k}\int_{RR; z>z'}\chi_R'(k,z)(L-z)\chi_R'(-k,z')\nonumber\\
&&-\frac{\pi}{k^3}\int_z \chi_L(k,z)\chi_L(-k,z)-\frac{2\pi}{k}\int_{LL; z>z'}\chi_L(k,z)(L+z')\chi_L(-k,z')\nonumber\\
&&-\frac{\pi}{k^3}\int_z \chi_L'(k,z)\chi_L'(-k,z)-\frac{2\pi}{k}\int_{LL; z>z'}\chi_L'(k,z)(L+z')\chi_L'(-k,z')\nonumber\\
&&
+ \frac{\pi}{2kL}\left(1+\frac{\pi}{2kL}\left(kd-2coth(kd)\right)\right)(c_R^2+c_L^2)
-\frac{\pi}{k^2L^2}csch(kd)c_Lc_R\Bigg\}
\end{eqnarray}

This expression is quite different from the analogous expression for the transverse field. Most notably the entanglement is  due to the last term in eq.(\ref{rhos}), which entangles the nonlocal modes of $\chi$. To calculate the entropy we therefore follow a similar root to \cite{first} and decompose the field in the basis which in particular contains the integrals of $\chi_L$ and $\chi_R$ over the corresponding space region. In other words we write

\begin{align}
\begin{split}
&
\chi^L(k,z)=\chi^L_0(k)+\tilde{\chi}^L(k,z)  \ \ \ \  \  \ : \ \ \ \ \ \ \chi^L_0(k)=\frac{1}{L-d/2}\int_L\chi^L(k,z') \ \ \ \ \ \ \ \ \ \ \ 0=\int_L\tilde{\chi}^L(k,z')
\\&
\chi^R(k,z)=\chi^R_0(k)+\tilde{\chi}^R(k,z)  \ \ \ \  \  \ : \ \ \ \ \ \ \chi^R_0(k)=\frac{1}{L-d/2}\int_R\chi^R(k,z') \ \  \ \ \ \ \ \ \ \ 0=\int_R\tilde{\chi}^R(k,z')
\end{split}
\end{align}
The fields $\tilde\chi_L$ and $\tilde\chi_R$ can themselves be decomposed in the basis of functions orthogonal to $\chi_{L,R}^0$. We will not pursue the exact form of such basis functions. The reason is that we are interested only in calculating the contribution to the entropy due the nonlocal mode $\chi_0$. Thus although one does expect a contribution due to the modes  $\tilde \chi$ or their mixing with $\chi_0$ which will be localized on the boundary of the slab (just like the contribution from $\zeta$ in the transverse case), we will not study these  contributions. Therefore the exact form of the basis for $\tilde\chi$
is of no importance to us. The terms in the density matrix that involve the interesting modes are (discarding the terms that vanish in the limit $L/d\rightarrow \infty$, $Lk\rightarrow\infty$)

\begin{align}\label{dens5}
\begin{split}
\rho[\chi^L_0,\chi^{L'}_0,\chi^{R}_0,\chi^{R'}_0]
&
\approx exp\Bigg\{-\frac{\pi L^3}{3k}[\chi^L_0(k)\chi^L_0(-k)+\chi^{L'}_0(k)\chi^{L'}_0(-k)]
\\&
-\frac{\pi L^3}{3k}[\chi^R_0(k)\chi^{R}_0(-k)+\chi^{R'}_0(k)\chi^{R'}_0(-k)]
\\&
+ \frac{\pi L^3}{8k}\left(1+\frac{\pi}{2kL}\left(kd-2coth(kd)\right)\right)\left([\chi^L_0(k)+\chi^{L'}_0(k)]^2+[\chi^R_0(k)+\chi^{R'}_0(k)]^2\right)
\\&
-\frac{\pi L^2}{4k^2}csch(kd)\left( [\chi^L_0(k)+\chi^{L'}_0(k)][\chi^R_0(k)+\chi^{R'}_0(k)] \right)
\Bigg\}
\end{split}
\end{align}
Or
\begin{equation}
\rho[\chi,\chi']=N\exp\Big\{-\frac{\pi L^3}{2k}\Big[\chi X\chi+\chi' X\chi'-2\chi Y\chi'\Big]\Big\}
\end{equation}
where $\chi\equiv (\chi_L,\chi_R)$, $\chi'\equiv (\chi_L',\chi_R')$ and

\begin{align}
X=\frac{2}{3}-Y
\ \ \ \ \ \ \ \ \ \ \ \ \ 
Y=\left(
\begin{array}{cc}
 \frac{d k-2 \coth (d k)+2 k l}{8 k l} & -\frac{\text{csch}(d k)}{4 k l} \\
 -\frac{\text{csch}(d k)}{4 k l} & \frac{d k-2 \coth (d k)+2 k l}{8 k l} \\
\end{array}
\right)
\end{align}
Or
\begin{equation}
\rho[\chi,\chi']=N\exp\Big\{-\frac{\pi L^3}{2k}\Big[\chi (\mu+\lambda)\chi+\chi' (\mu+\lambda)\chi'-2\chi (-\mu+\lambda)\chi'\Big]\Big\}
\end{equation}
with

\begin{align}
\mu=\left(
\begin{array}{cc}
 \frac{-3 d k+6 \coth (d k)+2 k l}{24 k l} & \frac{\text{csch}(d k)}{4 k l} \\
 \frac{\text{csch}(d k)}{4 k l} & \frac{-3 d k+6 \coth (d k)+2 k l}{24 k l} \\
\end{array}
\right)
\ \ \ \ \ \ \ \ \ \ \ \ \ 
\lambda=
\frac{1}{3}\begin{pmatrix}
1 & 0\\
0& 1
\end{pmatrix}
\end{align}

To calculate entropy we need the eigenvalues of $\mu^{-1}\lambda$, which are

\begin{equation}
\begin{split}
&
\lambda_1=
4
+\frac{6 \left(dk-2 \coth \left(\frac{dk}{2}\right)\right)}{kL}
+\frac{9 \left(d k-2 \coth \left(\frac{d k}{2}\right)\right)^2}{k^2 L^2}
+O(L^{-3}) 
\\&
\lambda_2=
4
+\frac{6 \left(dk-2 \tanh \left(\frac{dk}{2}\right)\right)}{kL}
+\frac{9 \left(d k-2 \tanh \left(\frac{d k}{2}\right)\right)^2}{k^2 L^2}
+O(L^{-3}) 
\end{split}
\end{equation}
We use

\begin{align}\label{sl}
\sigma^{L}=\frac{1}{2}tr\left\{   ln\left[\frac{\mu^{-1}\lambda-1}{4}\right] +\sqrt{\mu^{-1}\lambda} \ \cosh^{-1}\left[\frac{\mu^{-1}\lambda+1}{\mu^{-1}\lambda-1} \right]   \right\}
\end{align}
Keeping only the leading contribution and the first exponential correction we obtain\footnote{Eq.(\ref{sl}) also yields terms which are of order $d/L$ and have no exponential suppression factor.  However we have neglected such terms while calculating  $c_R$ and $c_L$. We have therefore not included these in Eq.(\ref{slk}).}
\begin{equation}\label{slk}
\sigma^L=
\log \left(\frac{27}{4}\right)
-\frac{6 \ln 3}{kL}e^{-2kd} 
\end{equation}

 Integrating  over all transverse momentum modes we finally obtain \begin{equation}\label{sl}
S=\frac{L_\perp^2}{a^2}\left[\log \left(\frac{27}{4}\right)
-\frac{6 \ln 3}{kL}e^{-2kd} \right]
\end{equation}
Where $\frac{2\pi}{a}$ is the ultraviolet cutoff on the transverse momentum integration and $L_\perp$ is the size of the system in the transverse direction.

The first term like for the transverse contribution is equal to twice the entropy of the single boundary case.
The correction is negative, as expected and decays with the same exponential factor as in the transverse case. The surprising feature of eq.(\ref{slk}) is that the correction is suppressed by a factor $1/kL$.
Thus the contribution of the nonlocal mode is "superlocal" - no correction due to the final width of the slab is present in the thermodynamic limit\footnote{We still expect finite correction proportional to $\exp\{-2kd\}$ coming from $\tilde\chi$, but this is a different issue.}.

\section{Conclusions}
In this note we have calculated the entanglement entropy of pure Maxwell fields in the  slab geometry. Our main interest was in the contribution of the (2D) longitudinal nonlocal mode of the magnetic field. Our main result is eqs.(\ref{slk}, \ref{sl}). The striking result is that even though the entangled field modes are nonlocal (integrals of $\chi$ over the two semi infinite regions of space), the contribution to the entropy  does not betray any signs of nonlocality. It  is consistent with the entanglement entropy that originates mostly in the boundary region (the first term in eq.(\ref{slk}), with a correction which exponentially decreases with the width of the slab as $e^{-2kd}$.  This type of behavior is intuitively expected in a local theory where the range of entanglement is of the same order as the correlation length.  

Recall that to set up our calculation in terms of independent degrees of freedom, we have expressed the third component of magnetic field $B_3$ via the components of magnetic field parallel to the slab surface. The resulting expression for $B_3$ in terms of $\chi$  has apparent nonlocality. Nevertheless in physical terms $B_3$ is a local field, i.e. its canonical commutation relations with the energy density and electric field are local, and thus it would be natural that its effects on all physical observables including entanglement entropy should be like that of a local field. 

Indeed consider the definition eq.(\ref{b3}). If one assumes that fields decay at infinity, i.e. $B_3(z\rightarrow\pm\infty)\rightarrow 0$, this definition implies 
\begin{equation}
\chi_0^L(k)=B_3(k,z=-d/2);\ \ \ \ \ \chi_0^R(k)=B_3(k, z=d/2)
\end{equation}
Thus the entanglement arising in the longitudinal sector is precisely due to the entanglement of the modes of $B_3$ at the boundary of the slab\footnote{
The exponentially decaying correction to the entanglement entropy of two disconnected boundaries must have an  interpretation as entanglement due to this physically local degree of freedom with finite correlation length.
In this respect our result in eq.(\ref{slk}) is somewhat peculiar in that the exponential is multiplied by a prefactor $1/kL$, which vanishes in the thermodynamic limit. It thus appears that in infinite volume there is no communication through $B_3$ between the two boundaries of the slab, at least as long as $kd\gg1$. However this conclusion is not very robust. Recall that when calculating the entropy we have neglected any contribution from the local  longitudinal modes $\tilde\chi$ of as well as their mixing with $\chi_0$. Such contribution can certainly bring about an exponential correction which is unsuppressed by an extra power of inverse size of the system.}.

\bibliography{ref}

\end{document}